\documentclass{article}

\usepackage{subfigure}
\usepackage{arxiv}
\usepackage{amsmath}
\usepackage{lineno}
\usepackage{graphicx}
\usepackage{svg}
\usepackage[utf8]{inputenc} 
\usepackage[T1]{fontenc}    
\usepackage{hyperref}       
\usepackage{url}            
\usepackage{booktabs}       
\usepackage{amsfonts}       
\usepackage{nicefrac}       
\usepackage{microtype}      
\usepackage{graphicx}

\title{SuperdropNet: a Stable and Accurate Machine Learning Proxy for Droplet-based Cloud Microphysics}

\author{{Shivani Sharma}\thanks{Corresponding author} \\
	Model-Driven Machine Learning\\
        Institute of Coastal Systems - Analysis and Modeling\\
        Helmholtz-Zentrum Hereon\\
	Germany \\
	\texttt{shivani.sharma@hereon.de} \\
	\And
	{David Greenberg} \\
	Model-Driven Machine Learning\\
        Institute of Coastal Systems - Analysis and Modeling\\
        Helmholtz-Zentrum Hereon\\
	Germany \\
	\texttt{david.greenberg@hereon.de} \\
}

\renewcommand{\shorttitle}{SuperdropNet: Machine Learning Parameterization for Superdroplet\\ 
 Cloud Microphysics Scheme}

\begin{document}
\maketitle

\begin{abstract}
	Cloud microphysics has important consequences for climate and weather phenomena, and inaccurate representations can limit forecast accuracy. While atmospheric models increasingly resolve storms and clouds, the accuracy of the underlying microphysics remains limited by computationally expedient bulk moment schemes based on simplifying assumptions. Droplet-based Lagrangian schemes are more accurate but are underutilized due to their large computational overhead. Machine learning (ML) based schemes can bridge this gap by learning from vast droplet-based simulation datasets, but have so far struggled to match the accuracy and stability of bulk moment schemes. To address this challenge, we developed SuperdropNet, an ML-based emulator of the Lagrangian superdroplet simulations. To improve accuracy and stability, we employ multi-step autoregressive prediction during training, impose physical constraints, and carefully control stochasticity in the training data. Superdropnet predicted hydrometeor states and cloud-to-rain transition times more accurately than  previous ML emulators, and matched or outperformed bulk moment schemes in many cases. We further carried out detailed analyses to reveal how multistep autoregressive training improves performance, and how the performance of SuperdropNet and other microphysical schemes hydrometeors' mass, number and size distribution. Together our results suggest that ML models can effectively emulate cloud microphysics, in a manner consistent with droplet-based simulations.
\end{abstract}


\section{Introduction}
In early versions of weather and climate models, low spatial resolution were the primary source of model errors \cite{manabe_climate_1969}. Over time, finer grids and shorter time steps improved accuracy by explicitly representing processes that would otherwise occur at sub-grid scales, but they impose huge memory and computing requirements that push against the limits of available hardware \cite{palmer_vision_2020}.

For operational weather forecasting and long term climate projections, it is still standard practise to parameterize sub-grid scale processes  \cite{gross_physicsdynamics_2018,palmer_vision_2020} such as convection, radiation and cloud microphysics. All parameterization schemes involve some form of approximation which, while speeding up calculations, lead to errors and uncertainties in simulations and forecasts \cite{gross_physicsdynamics_2018}. In numerical weather prediction (NWP) models, systemic errors accumulate over time and after forecast lead times of 7-10 days model errors tend to be of the same magnitude as the predicted signals \cite{palmer_vision_2020}. parameterizations contribute significantly to these model errors \cite{gross_physicsdynamics_2018,palmer_vision_2020}.

In recent years, a data driven approach has been applied in many instances to replace the traditional parameterization schemes. Essentially, simulation routines too computationally intensive for use in operational forecasting are used to generate training data for optimizing a machine learning (ML) model. This approach has been applied to convection \cite{brenowitz_prognostic_2018, gentine_could_2018, ogorman_using_2018, rasp_deep_2018,yuval_use_2021}, radiative transfer \cite{veerman_predicting_2021, belochitski_robustness_2021} gravity wave effects \cite{chantry_machine_2021, dong_accelerating_2023}, atmospheric chemistry \cite{kelp_toward_2020, kelp_online-learned_2022} and turbulence \cite{leufen_calculating_2019}. Many ML models perform well in predicting single time steps but exhibit instability when run for many time steps during online testing. The primary source of instability here is the accumulation of error over longer integration times. 

Here we focus on the task of parameterizing cloud microphysics, and specifically the coalescence of liquid cloud droplets into rain. Weather prediction models typically employ bulk moment schemes \cite{seifert_double-moment_2001} that simplify particle size distributions into the total mass and number densities of droplets above and below a chosen cloud/rain threshold size, and rely on approximate physics and statistical assumptions to update these quantities over time. While bulk moment schemes are fairly consistent with droplet-based simulations in simplified scenarios, they struggle in the presence of mixed-phase clouds, particularly in the representation of ice microphysical processes\cite{khain_representation_2015,morrison_confronting_2020}. Ultimately, inaccurate cloud microphysics schemes manifest as inaccurate precipitation forecasts \cite{lynn_utilization_2007} and errors in radiative transfer calculations.

In \cite{gettelman_machine_2021} an ML emulator for a bin microphysics scheme was developed and coupled to a general circulation model for an improved representation of cloud and rain particle distributions in a warm rain scenario. Another study showed that neural networks can be trained for computing bulk moment dynamics to match superdroplet simulations \cite{seifert_potential_2020}. This approach combines the low memory and compute requirements of bulk moment schemes with the accuracy, simplicity and physical consistency of superdroplet simulations. It also offers straightforward compatibility with existing atmospheric models that rely on bulk moment representations to simulate radiation transfer, convection and other moisture-dependent processes. However, while the networks accurately predicted instantaneous rates of various coalescence events, they were far less accurate at predicting the evolution of bulk moments over the longer time scales of cloud-to-rain transitions. Thus, ML currently exhibits a major performance gap compared to classical bulk moment parameterizations.
 
To address this challenge we developed SuperdropNet, an emulator for superdroplet simulations in a warm rain scenario. We introduce several innovations in designing and training our network that improve accuracy and stability over long time integration windows for diverse initial conditions. These include autoregressive prediction of multiple time steps during training \cite{grzeszczuk_neuroanimator_1998,um_solver---loop_2020,kelp_toward_2020, kelp_online-learned_2022}, mass conservation as a hard constraint, relaxing some assumptions of bulk moment schemes and carefully controlling stochasticity when generating training data. We also systematically analyze how the length of autoregressive rollouts during training affect forecast accuracy at various time horizons.  We compare SuperdropNet to a traditional warm rain bulk moment scheme \cite{seifert_double-moment_2001} commonly used in the ICON (Icosahedral Nonhydrostatic) model, and to a previously described ML-based parameterization \cite{seifert_potential_2020}. Our results show that SuperdropNet significantly closes the accuracy gap between ML parameterizations and bulk moment schemes, and even outperforms classical schemes for some initial conditions. We further examine performance of these schemes, and identify how their accuracy can depend on various factors in the simulated scenario. 

\section{Warm rain microphysical simulations}
\subsection{Droplet schemes}
Simulating droplets in a numerical simulation provides the most accurate estimation of droplet interactions that contribute to the cloud microphysical processes. However, individually simulating droplets can be computationally expensive even for small domain sizes. In \cite{shima_super-droplet_2009} this problem is simplified by using `superdroplets' to represent multiple individual droplets of same size that are close to each other. The motion of the droplets is simulated along with collision-coalescence, condensation/evaporation and sedimentation processes. This method uses a Monte Carlo scheme for estimating the collision-coalescence process. The droplet size distribution is initially assumed to be gamma, exponential or a log-normal \cite{marshall_distribution_1948}, and is evolved in time by sampling pairs of colliding droplets from it.


\subsection{Bulk moment schemes}
\label{bulk-momemts}
In most atmospheric models, cloud microphysical processes are represented using bulk moment schemes. Instead of simulating droplets and tracking collision probabilities, bulk moment schemes track only the evolution of the first and sometimes the zeroth moment of the droplet distribution. A warm rain scenario, with only clouds and rain present, is fully described by a density function $f(x)$ over droplets with mass $x$. This density function is used to define 4 bulk moments:
\begin{align}
    L_{c} = \int_{0}^{x^{*}} {x}f(x)dx & &
    N_{c} = \int_{0}^{x^{*}} f(x)dx \nonumber\\
    L_{r} = \int_{x^{*}}^{\infty} {x}f(x)dx & &
    N_{r} = \int_{x^{*}}^{\infty} f(x)dx
\end{align}
Here $x^*=2.6 {\times} 10^{-10}$kg is the mass threshold dividing cloud and rain and droplets \cite{beheng_general_1986}. The zeroth moments, $N_c$ and $N_r$, are the number density of cloud and rain droplets respectively. The first moments, $L_c$ and $L_r$, represent the total cloud and rain water mass. 

To provide a baseline when evaluating ML-based microphysical schemes, we employ here the two-moment bulk scheme of \cite{seifert_double-moment_2001}. The time evolution of cloud water, rain water, cloud droplet concentration and rain droplet concentration is estimated through the ordinary differential equation system given by:
\begin{align}
\frac{dL_c}{dt} &= -AU -AC \\
\frac{dL_r}{dt} &= +AU +AC \\
\frac{dN_c}{dt} &= -2AU_n - AC_n - SC_c = \frac{-2}{ x^*}AU -\frac{1}{\overline{x_c}} AC - SC_c \\
\frac{dN_r}{dt} &= +AU_n + AC_n - SC_r = \frac{1}{ x^*}AU  - SC_r 
\end{align}
The autoconversion rate $AU$ is the rate at which cloud mass converts to rain mass due to collisions between the cloud droplets, while the accretion rate $AC$ describes the mass flux associated with collisions between rain and cloud droplets. The mean cloud droplet mass ${\overline{x_c}=L_{c}/N_{c}}$ is the average mass of cloud droplets. $AU_n$ and $AC_n$ are autoconversion and accretion rates corresponding to the number concentrations and are calculated by assuming that autoconversion events involve droplets with an average mass of $x^*$ and that accretion events lead to the formation of cloud droplets with an average mass ${\overline{x_c}}$.  Self collection rates $SC_{c}, SC_{r}$ describe the rate of collisions that do not convert cloud droplets to rain.

Following standard practice for cloud microphysics in atmospheric models such as ICON (\cite{zangl_icon_2015}), we carry out explicit (Euler) integration of these differential equations with fixed time step $\Delta t$. In practice, $\Delta t$ used for microphysical processes may be shorter than the time step used for the dynamical core in an atmospheric model, and can vary from a few minutes to a few seconds. 

\cite{seifert_double-moment_2001} employs several approximations and statistical assumptions to derive formulas for calculating the process rates in a warm rain scenario. This approach of approximating process rates to compute changes in bulk moments is almost universally applied in operational weather and climate models and formed the basis for the ML approach formulated in \cite{seifert_potential_2020}.

\section{Problem statement}
\label{sec:problem}
This study aims to develop a scheme for warm-rain microphysics that efficiently computes bulk moment dynamics consistent with superdroplet simulations. Having described droplet-based and bulk moment schemes, we can now give a concrete formal description of this task. We use $y_t$ to denote the vector of bulk moments at time $t$ in a superdroplet simulation.
\begin{align}
    y_t &= [L_c(t),\quad L_r(t),\quad N_c(t),\quad N_r(t)]
\end{align}
A classical or ML-based scheme is defined by a time-stepping function $\mathcal M$, which can be applied to $y_t$ to compute $\mathcal M(y_t) \approx y_{t+1}$. We can also iteratively apply $\mathcal M$ $k$ times, feeding the outputs back in as inputs to generate an autoregressive prediction. We denote this repeated application by $\mathcal M^{(k)}$, and can use to predict $y$, $k$ steps into the future.
\begin{align}
    \mathcal M^{(k)}(y_t) &= \overbrace{\mathcal M \circ \mathcal M \circ \cdots \circ \mathcal M}^{k \text{ times}}(y_t) \approx y_{t+k}
\end{align}
We refer to sequence of bulk moment vectors $\left\{\mathcal M (y_t), \mathcal M^{(2)} (y_t), \ldots, \mathcal M^{(k)} (y_t) \right\}$ computed by $k$ repeated applications of $\mathcal M$ as a length-$k$ \textit{rollout}.

Our goal is then to obtain an $\mathcal M$ that can evolve bulk moments, starting from initial conditions $y_0$, to match the full course of a superdroplet simulation:
\begin{align}
    \mathcal M^{(t)}(y_0) &\approx y_t,\quad\forall t
\end{align}
After giving further details on the superdroplet simulations used for training and evaluation (section \ref{sec:datagen}), we will describe how we use neural networks to define $\mathcal M$, and how we design and minimize a loss function to achieve our stated aims (section \ref{sec:deeplearning}).

\section{Data Generation with Droplet Simulations}
\label{sec:datagen}

For generating the training data, we simulate superdroplets in a warm rain scenario in a zero-dimensional box. The superdroplet simulations were carried out using the McSnow software \cite{brdar_mcsnow_2018,seifert_potential_2020}, and the output is recorded every $\Delta t=20$ s. This time step was chosen to be consistent with previous work \cite{seifert_potential_2020}, and to fall within the typical range for atmospheric modeling.

Superdroplet simulations were used to compute bulk moments as follows:
\begin{align}
    L_{c} &= \sum_{\forall i: x_i<x^*} {x_c}{\xi_c} & &
    N_{c} = \sum_{\forall i: x_i<x^*} {\xi_c} \nonumber\\
    L_{r} &= \sum_{\forall i: x_i\geq x^*} {x}{\xi_r} & &
    N_{r} = \sum_{\forall i: x_i\geq x^*} {\xi_r}
\end{align}
Here mass moments ($L_c$, $L_r$) measure the total mass of superdroplets above and below the cloud-rain threshold, while number counts ($N_c$, $N_r$) count the total number to cloud and rain droplets.

\subsection{Initial conditions}
\begin{figure}[h]
\centering
\includegraphics[width=12cm,height=8cm,scale=0.5,trim={6cm 2cm 0cm 3cm},clip]{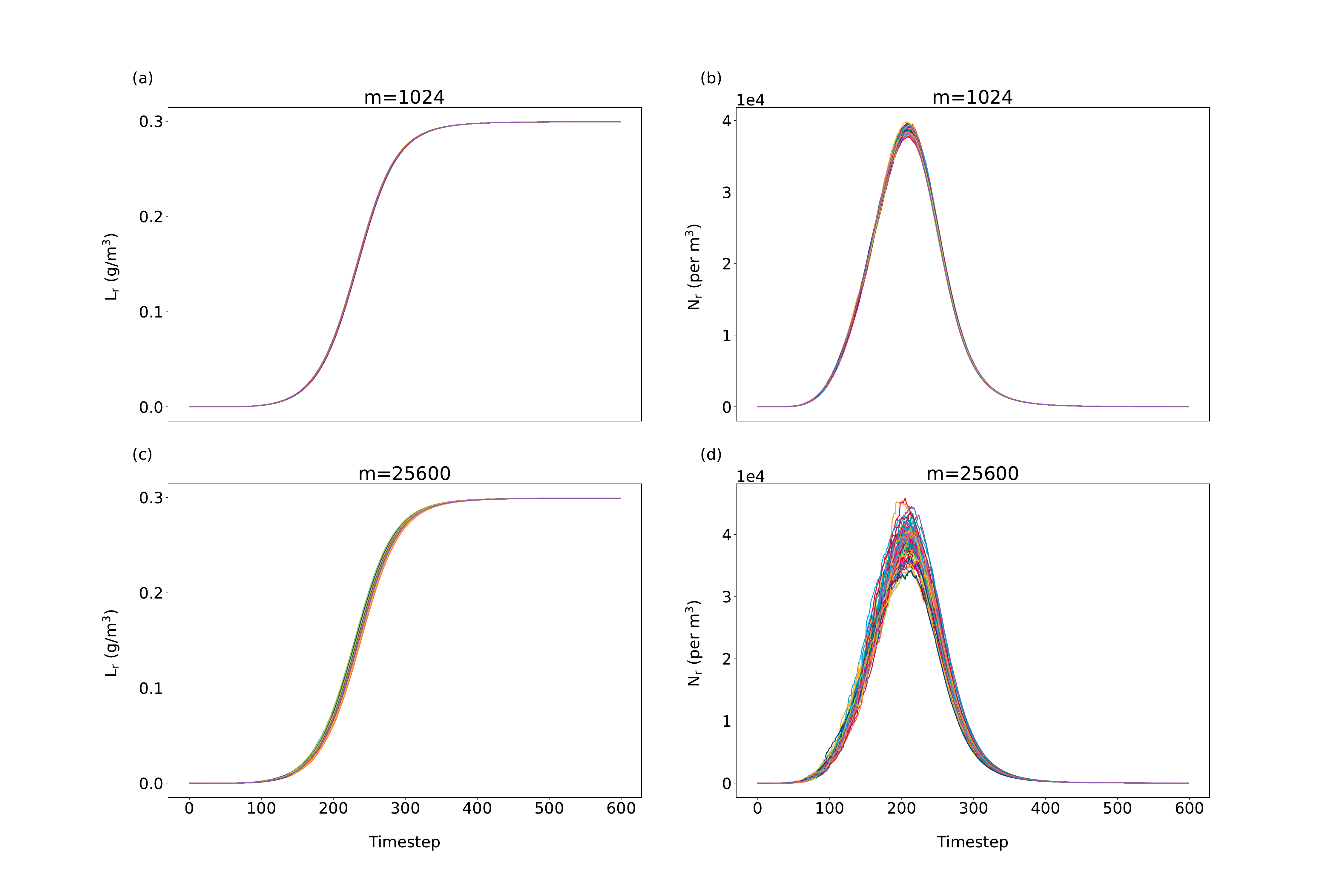}
\caption {Stochasticity of superdroplet simulations with box size $25m^{3}$. 
(a) and (b) correspond to 100 simulations at multiplicity of 1024. (c) and (d) correspond to 100 simulations at multiplicity of 25600. For both sets of simulations, rain water mass ($L_r$) and rain number concentration($N_r$) are compared. Other initial conditions for both simulations are the same with ${L_0}$=0.3 g/m$^3$, $r_0$=13 $\mu$m, $\nu$=0.5.}
\end{figure}

\begin{figure}[h]
\centering
\includegraphics[width=12cm,height=8cm,scale=0.5,trim={6cm 2cm 0 3cm},clip]{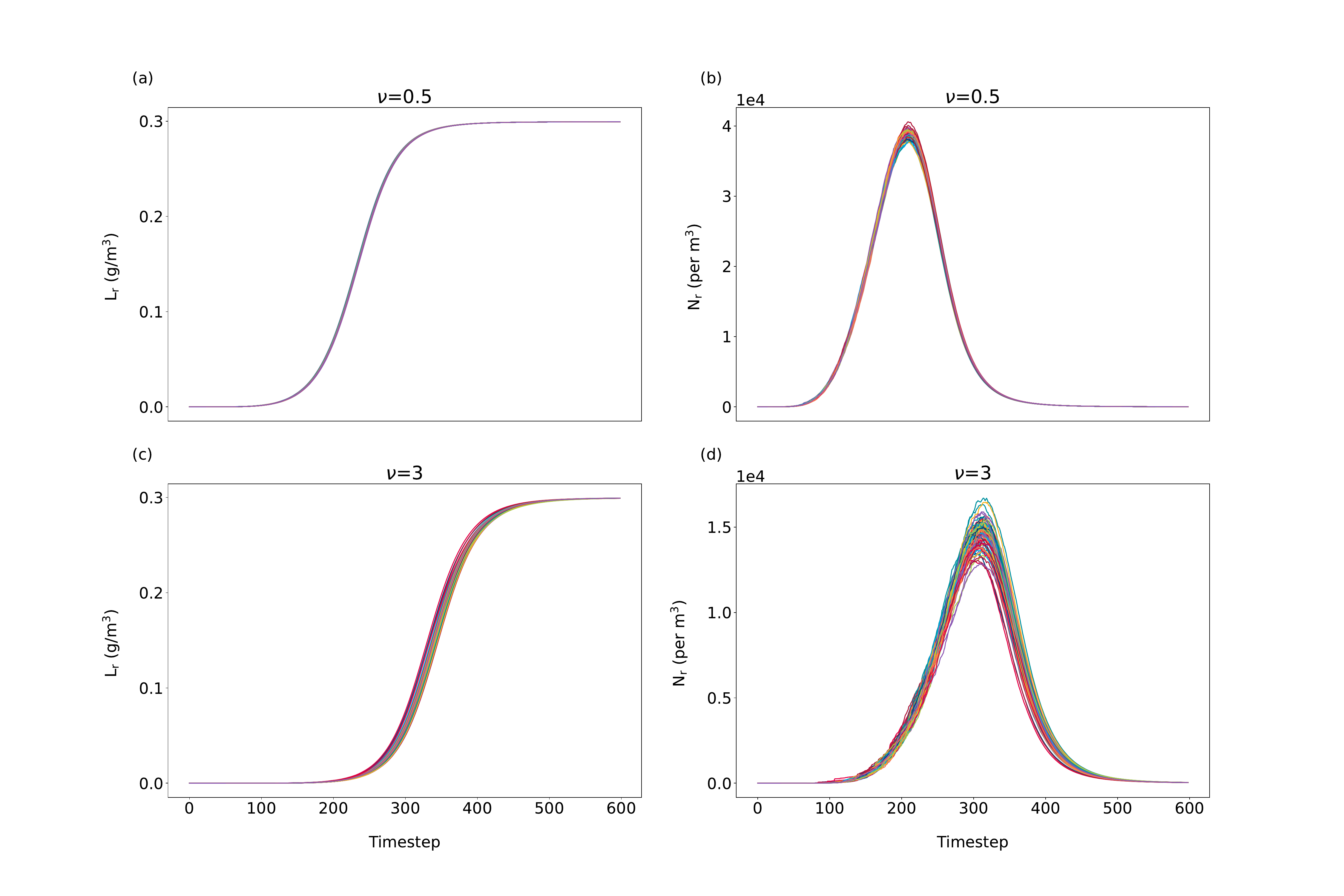}
\caption{Stochasticity in simulations at a box size of 2500 $m^{3}$ and multiplicity is 25600. (a) and (b) correspond to 100 simulations at $\nu$=0.5. (c) and (d) correspond to 100 simulations at $\nu$=3. For both sets of simulations, Rain water mass($L_r$) and Rain number concentration($N_r$) are compared. Other initial conditions for both simulations are the same with ${L_0}$=0.3 g/m$^3$, $r_0$=13 $\mu$m.}
\end{figure}
For all simulations the box volume and multiplicity are fixed to be 2500 m$^3$ and 26500, respectively. In the context of superdroplet simulations, multiplicity refers to the number of individual droplets represented by a single superdroplet. Superdroplet simulations are initialized by sampling droplets with total mass $L_0$ from a gamma distribution, with shape parameter $\nu$ and mean droplet radius $r_0$. Higher values of $\nu$ indicate a more peaked size distribution. We sample from a huge range of parameter values as described in Table~\ref{ic_table}. In total, these combinations of parameter values lead to 819 distributions.

\begin{table}[h]
 \caption{Initial Conditions for box-model simulations}
 \centering
 \begin{tabular}{l c}
 \hline
  Quantity  & Range  \\
 \hline
   $L_0$  &\{0.2,0.3,0.4,0.5,0.6,0.7,0.8,0.9,1.0,1.2,1.5,1.6,2.0\} g/m$^3$  \\
   $r_0$  & \{9, 10, 11, 12, 13, 14, 15\} $\mu$m \\
   $\nu$  & \{0, 0.5, 1, 1.5, 2, 2.5, 3, 3.5, 4\}  \\
 \hline
 \multicolumn{2}{l}{The complete set of ICs was the same as in \cite{seifert_potential_2020}} 
 \end{tabular}
 \label{ic_table}
\end{table}

\subsection{Measuring and controlling stochasticity}
As stated in section \ref{sec:problem}, our ultimate aim is to train a neural network to update bulk moments, using superdroplet data. A challenge is posed by the fact that superdroplet
simulations are inherently stochastic, and the same initial conditions can produce different results with in repeated simulations. In initial experiments, we found that randomness in training data, and in particular large infrequent jumps in the bulk moments, lead to overfitting of individual stochastic events, and instability in the optimization process used to train the network.

For fixed mean droplet radius and total water mass, randomness can be reduced by increasing the box volume or decreasing the multiplicity. since this increases the total number of simulated superdroplets. With more superdroplets, the effects of random collisions tend to better `average out'. However, computation time also increases quadratically in the number of droplets. Reducing the shape parameter of the initial distribution also limits stochasticity by reducing the relative contribution of extremely large or small droplets. 

We measured these effect of stochasticity by computing bulk moments from repetitive superdroplet simulation runs under the same set of initial conditions. We first examined the role of multiplicity using simulations wit a low volume (25 m$^3$), moderate den ($L_0=0.3$ g/m$^3$) and mean initial droplet size of 13 $\mu$m. For over 100 simulations with high multiplicity ($m=25600$), we observed considerable variation in the height and timing of the peak raindrop count(Fig 1-(d)), and noticeable variation in the time evolution of the rain droplet mass (Fig. 1-(c)). Simulations with a lower multiplicity $(m=1024)$ exhibited noticeably less variation in these quantities (Fig. 1-(a,b)), suggesting that this stochasticity is an artifact of the superdroplet technique and exceeds the stochasticity of the original droplet collision rules.

We next examined the effect of the shape parameter $\nu$ in additional simulations, using the same $L_0$ and $r_0$, multiplicity 25600, a larger box (2500 m$^{3}$) and $\nu$=0.5 or $3$. As previously observed \cite{seifert_potential_2020}, we found that higher $\nu$ values produced greater variation in bulk moments over repeated runs (Fig. 2). Similarly, variability was higher for smaller box sizes (Fig. 1, lower vs. Fig. 2, upper).

To limit stochasticity when generating training data for deep learning (see below), we used the larger box size of 2500 m$^3$. To limit computation time and allow better comparison with previous work, we kept the multiplicity value of 25600 from \cite{seifert_potential_2020}. To further limit variability, for each unique set of initial conditions ($r_0, \nu, L_0)$ we averaged bulk moments from 100 superdroplet simulations. We found that training on all individual simulations resulted in overfitting, poorer performance and longer computation times.

\subsection{Data preparation}
We carried out superdroplet simulations with 819 unique initial conditions (Table \ref{ic_table}). Bulk moments were calculated from the collection of superdroplets every 20 seconds, and averaged over 100 repeated simulations of each set of initial conditions. We randomly assigned 100 superdroplet simulations to testing and the remaining 719 were assigned for training and validation. From those 719 simulations, the data was randomly chunked and 90\% of it was assigned for training and 10\% for validation. We z-scored each dimension of the inputs to the neural network. 

\section{Deep learning of warm rain microphysics}
\label{sec:deeplearning}
\subsection{Time stepping with learned moment updates}
\label{comparison}
We chose to train neural networks to directly estimate the bulk moment tendencies $(y_{t+1}-y_t)/\Delta t$, so that the learned time stepping function is 
\begin{align}
\mathcal M(y_t) &= y_t + h_\theta(y_t, \phi)\Delta t  \approx y_{t+1}
\end{align}
$\theta$ are trainable parameters of the network, $h_\theta$ its input-output function and $\phi$ are additional inputs (details in sec. \ref{sec:net}).

This contrasts with the strategy presented in \cite{seifert_potential_2020} which instead estimates process rates, then uses the same approximations as the bulk moment scheme to compute droplet density-based process rates (sec. \ref{bulk-momemts}). We compare our results to one such network and refer to it as PRNet. Our approach can potentially provide a closer match to superdroplet schemes by avoiding these approximations. 

\subsection {Physically constrained deep learning}
To improve accuracy and physical consistency, we enforced constraints that hold in superdroplet and bulk moment simulations on our neural networks.
\begin{itemize}
    \item \textbf{Mass conservation} To ensure the total mass of rain and cloud droplets is conserved, we trained the neural network to predict cloud mass updates $ \Delta \widehat {L_c} = L_c(t+\Delta t) - L_c(t)$, and defined $ \widehat {L_r} = L_0 - \hat L_c$.
    \item \textbf{Irreversibility} In superdroplet and bulk moment simulations the total mass and count of cloud droplets can only decrease over time. To enforce this, we set any positive updates to cloud mass and droplet counts to zero at each time step (before calculating $\widehat{\Delta L_r}$). Initial experiments showed that this constraint interfered with learning by preventing backpropagation of loss gradients, so we used it only with trained models.
    \item \textbf{Positivity} Mass and droplet counts cannot be negative. We enforced this in a postprocessing step after moment prediction for all time steps, with negative moments set to zero while maintaining mass conservation. We did not use this constraint during training.
\end{itemize}

\subsection {Autoregressive Multi-step Training}
\label{ssec:multistep}
Our overall aim (sec. \ref{sec:problem}) is to predict the evolution of bulk moments through repeated application of a trained network. Clearly, if $\mathcal M(y_t) = y_{t+1}$ precisely for all $t$, then also $\mathcal M^{(j)}(y_t) = y_{t+j}$ for all $t$ and $j$. This suggests a simple `offline training' strategy of minimizing 1-step prediction error $\mathcal L_1 = \left\lVert y_{t+1} - \hat y_{t+1} \right\rVert$.

However, this fails in practice since $\mathcal L_1$ does not reflect the rate at which errors grow over a rollout. Thus when a network trained offline generates a multistep rollout, it inevitably makes at least some small errors and encounters inputs outside the training set. This can lead to inaccurate results or divergence to infinity as rollout length increases. This problem is particularly severe for simulations with lower water content, which require a greater number of time steps to produce rain. The inadequacy of offline training has been noted for several other parameterization tasks \cite{um_solver---loop_2020, kochkov_machine_2021, kelp_toward_2020, kelp_online-learned_2022}.

To address this limitation, we predict $k$ future time steps during training. Starting with superdroplet-derived bulk moments $y_t$, we use our network iteratively $k$ times: 
\begin{align}
\label{eq:loss}
\mathcal L_k &= \sum_{j=1}^k \lambda_j \left \lVert y_{t+j} - \mathcal M^{(j)}(y_t) \right \rVert
\end{align}
The loss is calculated over the moments as predicted by the iterative application of $g$ and the moments as calculated from the superdroplet simulations at the same time step. The scalar weights $\lambda_j$ allow emphasis of different prediction horizons during the training process. Initial experiments showed best results when setting $\lambda_k=1$ and all other $\lambda_0$. This choice of $\lambda_j$, known as the `pushforward trick,' has been previously observed to improve the performance of neural PDE solvers \cite{brandstetter_message_2022}, though the involved mechanism remains an open question.

\subsubsection{Increasing rollout length during training}
A challenge for optimizing functions such as $\mathcal L_k$, with repeated self-iteration of neural network, is that gradients can vanish towards zero or explode towards infinity over repeated iterations \cite{Goodfellow-et-al-2016}. This problem is particularly severe for networks at the start or early phase of training, since these tend to produce large errors and have not yet encountered inputs resembling their own erroneous outputs. To avoid instability during training, we begin in offline mode with $k=1$, and then gradually increase the value of $k$ during training. We train until convergence (sec. \ref{sec:opt}) on $\mathcal L_k$, then use the resulting network parameters to initialize training on $\mathcal L_{k+1}$, for $k\leq 25$.

While this procedure required longer to computation time than a single optimization procedure, we found that initializing the neural network from the weights of the previous $k$ value greatly reduced the number of optimization steps required. This is equivalent to a `warm start' in training as the neural network at $k+1$ receives pre-trained weights from $k$-th model instead of randomly initialized weights.

\subsection{Network Architecture}
\label{sec:net}
We use a fully-connected network with 3 hidden layers of 200 neurons each with ReLU (Rectified Linear Unit) activation functions. The total number of trainable network parameters was 83,200. In addition to $y_t$ the network receives 5 additional inputs $\phi$: the liquid water time scale ($\tau=L_r/L_0$), mean cloud droplet mass $\overline{x_c}=L_{c}/N_{c}$, $r_0$, $\nu$ and $L_0$. The outputs of the neural network are the tendencies for the four moments.

\subsection{Optimization Procedure}
\label{sec:opt}
We minimized $\mathcal L_k$ (eqn. \ref{eq:loss}) using the ADAM optimizer\cite{kingma_adam_2017} with initial learning rate 2e-4 for $k=1$ and the batch size as 256. As we subsequently increased $k$, the learning rate was decreased as the updates to the network weights also decreased in magnitude. The learning rate was halved if at the subsequent value of $k$, training ended before 10 epochs. This procedure to update the learning rate was followed until $k=24$. For $k=25$, learning rate was set to 2e-8 and decreasing it further did not yield continued reduction of the loss. 

Optimization for each $k$ used a maximum of 500 epochs. Training was cut short using early stopping when the validation loss did not decrease for 50 consecutive epochs. Bulk moments and network parameters were represented using 32-bit floating point.

\section{Results}

\subsection{Accuracy of Single Step Training}
\begin{figure}
\centering
\includegraphics[width=15cm,height=8cm,scale=0.5,trim={6cm 0 0 0},clip]{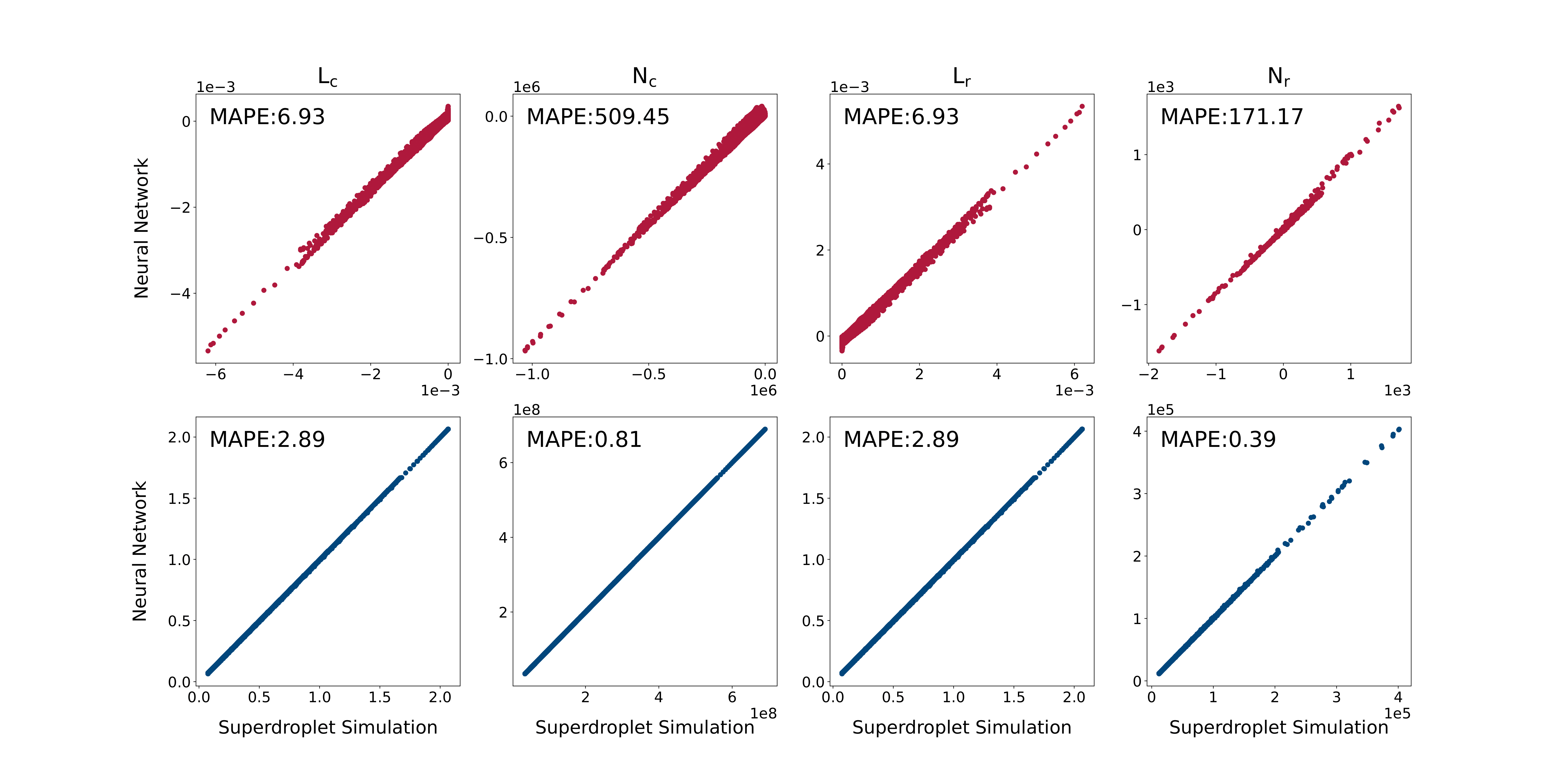}
\caption{Top panels, red -- Superdroplet-derived changes in bulk moments over single time steps ($\Delta t = 20$s) vs. changes over single time steps predicted by a neural network trained to predict one time steps into the future.
Bottom panels, blue -- Superdroplet-derived bulk moments vs. predictions on time step ahead by the same network. Results are shown only for initial conditions in the held-out testing data. Mean absolute percentage errors (MAPE) are shown for each comparison.}
\label{scatterplot1}
\end{figure}

\begin{figure}
\centering
\includegraphics[width=13cm,height=8cm,scale=0.5,trim={6cm 1cm 0 5cm},clip]{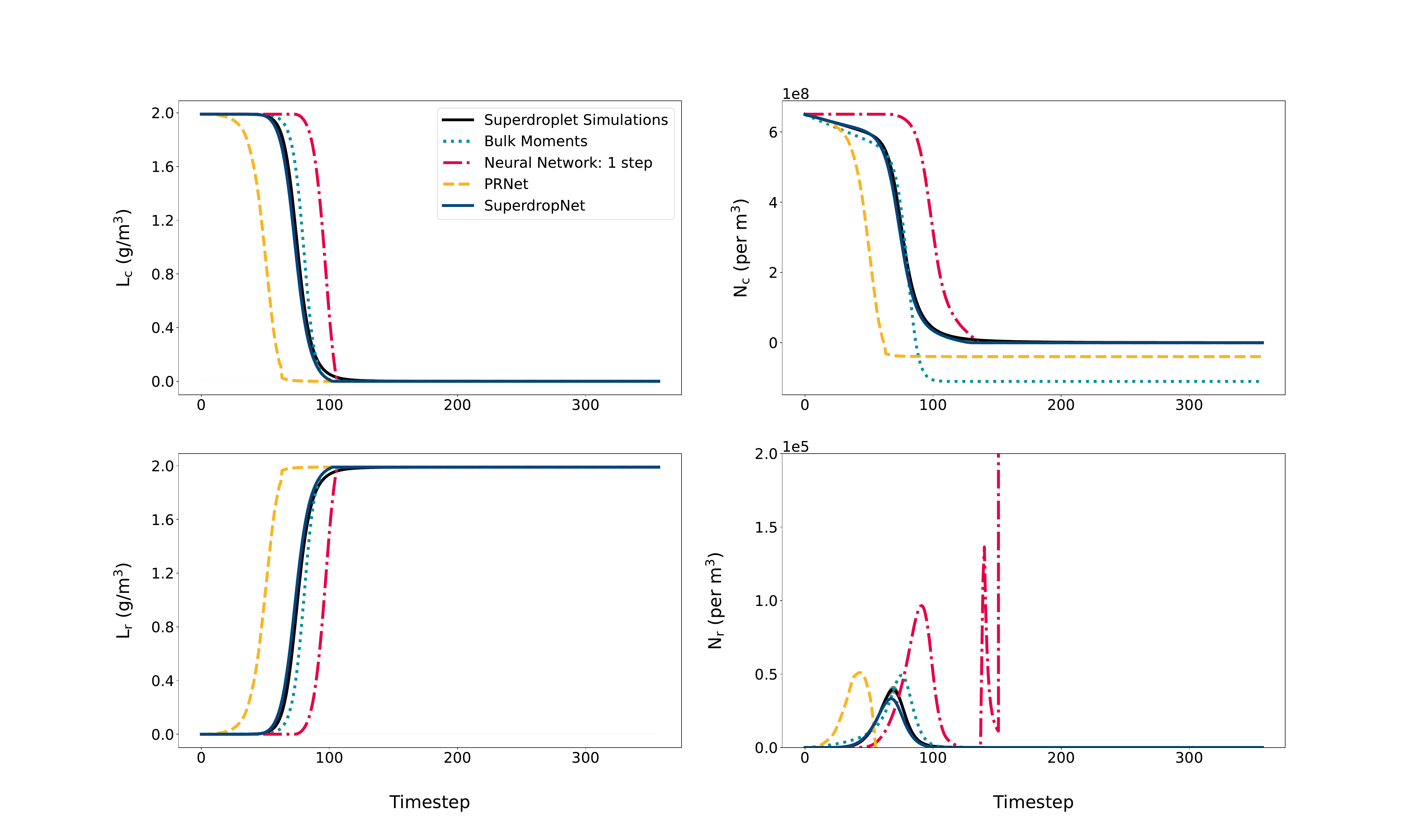}
\caption{Superdroplet-derived bulk moments (black lines) compared to rollouts from a neural network trained to predict 1 step into the future (red-dashed lines), SuperdropNet (blue solid line), PRNet (yellow-dashed lines) and from a classical bulk moment scheme (blue-dotted lines). Results are shown for a simulation with $L_0=2$ g/$m^3$, 
 $r_0=9\mu$m, $\nu=0$. Shaded region indicates +/- 1 standard deviation over 100 superdroplet simulations. A single time step corresponds to 20 s of simulation time.}
\label{ex_1}
\end{figure}

\begin{figure}
\centering
\includegraphics[width=13cm,height=8cm,scale=0.5,trim={6cm 1cm 0 5cm},clip]{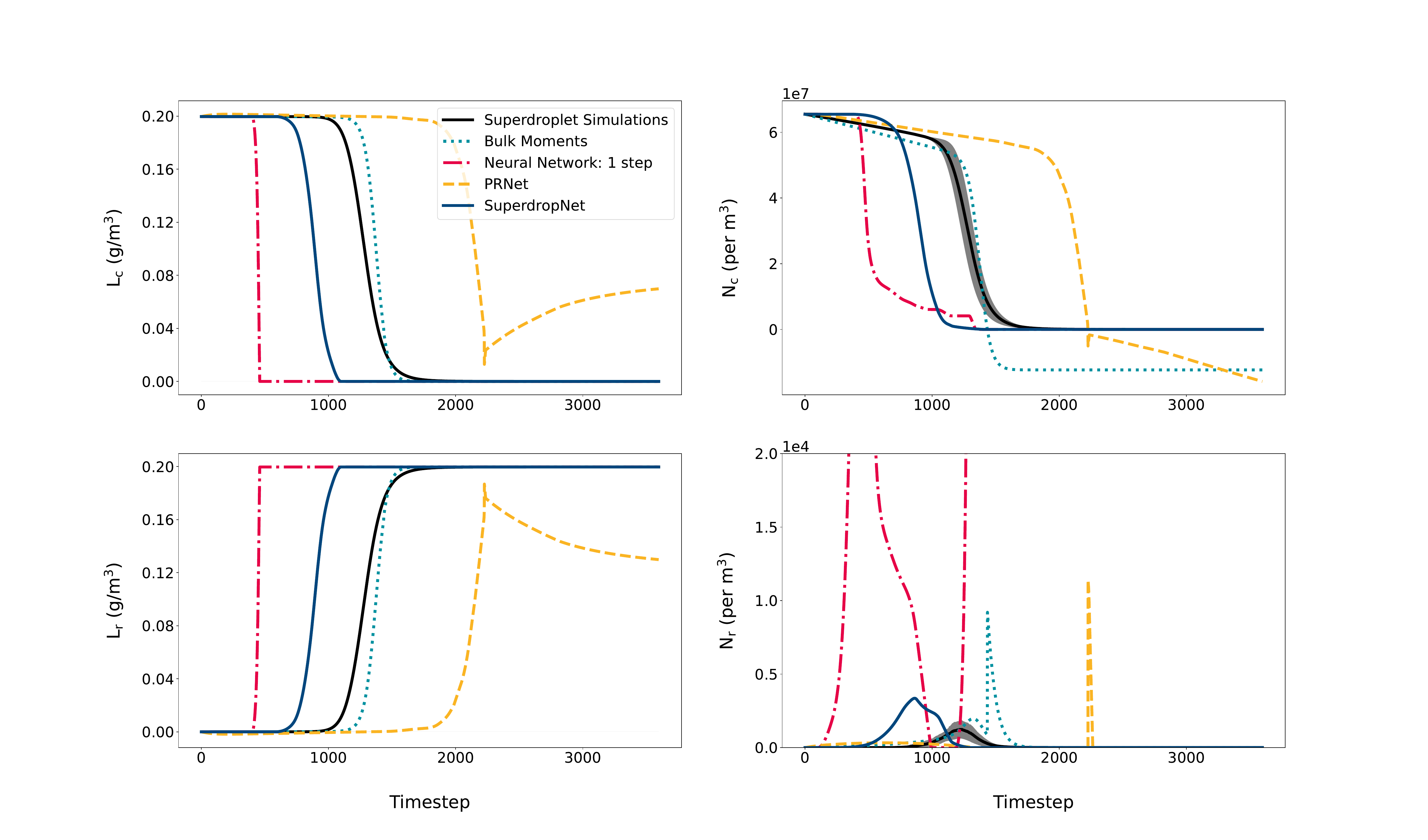}
\caption{Superdroplet-derived bulk moments (black lines) compared to rollouts from a neural network trained to predict 1 step into the future (red-dashed lines), SuperdropNet (blue solid line), PRNet (yellow-dashed lines) and from a classical bulk moment scheme (blue-dotted lines). Results are shown for a simulation with $L_0=0.2$ g/$m^3$, 
 $r_0=9\mu$m, $\nu=2$. Shaded region indicates +/- 1 standard deviation over 100 superdroplet simulations. A single time step corresponds to 20 s of simulation time.}
\label{ex_2}
\end{figure}

\begin{figure}
\centering
\includegraphics[width=13cm,height=8cm,scale=0.5,trim={6cm 1cm 0 5cm},clip]{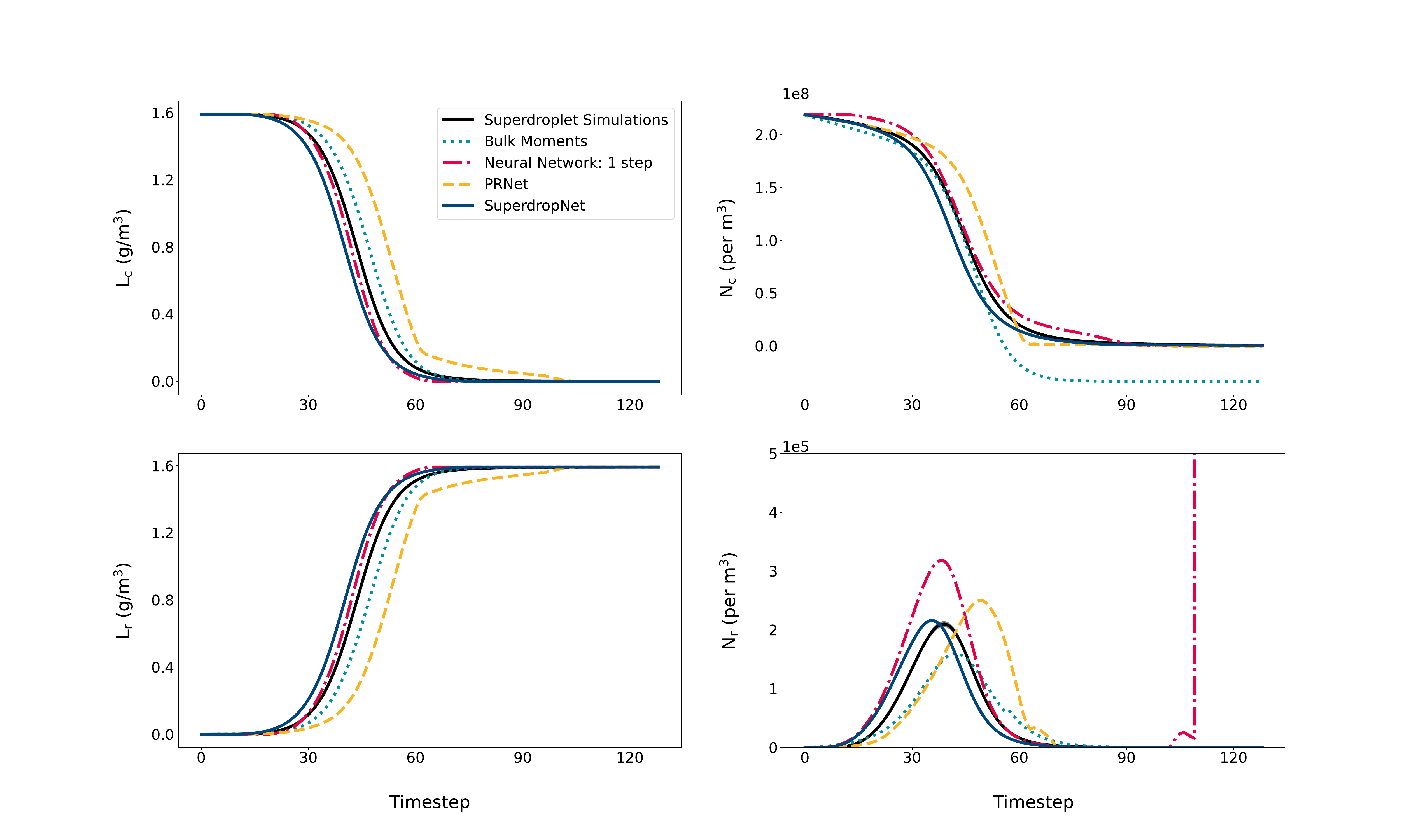}
\caption{Superdroplet-derived bulk moments (black lines) compared to rollouts from a neural network trained to predict 1 step into the future (red-dashed lines), SuperdropNet (blue solid line), PRNet (yellow-dashed lines) and from a classical bulk moment scheme (blue-dotted lines). Results are shown for a simulation with $L_0=1.6$ g/$m^3$, 
 $r_0=12\mu$m, $\nu=0$. Shaded region indicates +/- 1 standard deviation over 100 superdroplet simulations. A single time step corresponds to 20 s of simulation time.}
\label{ex_3}
\end{figure}

We first trained a neural network to predict superdroplet-derived bulk moments one time step into the future. After convergence, this `1-step' network closely predicted all 4 bulk moments one time step ahead (Fig. \ref{scatterplot1}-top panel). We calculated the mean absolute percentage errors(MAPE) between the predicted tendencies(direct output of the neural network) and superdroplet-derived bulk moment tendencies as well as between the predicted moments and the superdroplet-derived bulk moments. MAPE between the predictions $P$, of the actual values $A$,  is given by:
\begin{equation}
    {MAPE}(A, P) =  \frac{100}{N}\sum_{i=0}^{N} \frac{A_i - {P}_i}{A_i}
\end{equation}
where $N$ is the total number of samples.
Bulk moments at the next time step were inferred with mean absolute errors ranging from 0-3\% (Fig. \ref{scatterplot1}-bottom panel). Since $L_r$ and $L_c$ sum to $L_0$, their mean absolute percentage errors (MAPEs) are comparable, while MAPE is lower for droplet number concentrations. These results confirmed that our network architecture can closely predict how bulk moments will evolve over a single time step.

We next examined whether the 1-step network could predict bulk moments further into the future. Starting from the initial conditions of each simulation in our dataset, we iteratively applied the network to generate rollouts over the full course of the simulation. For example, in a simulation with high water content and an initial exponential droplet size distribution ($\nu=0$, Fig. \ref{ex_1}) the 1-step network (red-dashed line) roughly reproduced the bulk moments dynamics of superdroplet simulations (black line) except for $N_r$, where it diverged halfway through the simulation. 

We also examined a more challenging case (Fig. \ref{ex_2}), where low initial water content necessitated a longer rollout and $\nu=2$ produced a `wider' droplet size distribution. In this case, the 1-step network converted cloud to rain faster than superdroplet simulations. In a third case with intermediate water content, higher initial droplet radius and $\nu=0$ (Fig. \ref{ex_3}), the network matched superdroplet simulations more closely than the bulk moment parameterization for cloud and rain water content and cloud droplet concentration. For all three simulations (Fig. \ref{ex_1}-\ref{ex_3}), the 1-step network failed to produce stable and accurate predictions of rain droplet concentration.

\begin{figure}[h]
\centering
\includegraphics[width=15cm,height=8cm,scale=0.5,trim={6cm 0 0 0},clip]{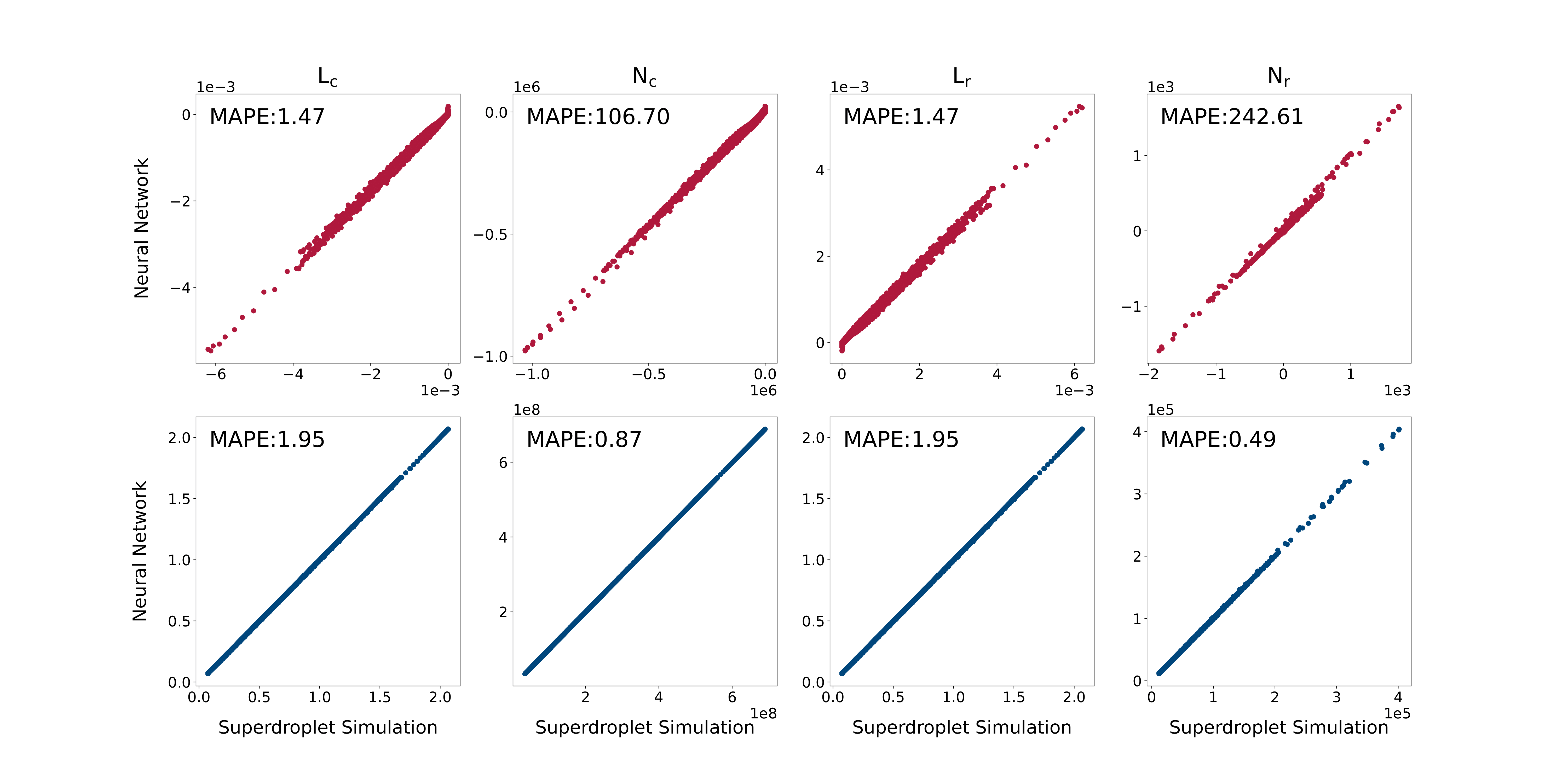}
\caption{Top panels, red -- Superdroplet-derived changes in bulk moments over single time steps ($\Delta t = 20$s) vs. changes over single time steps predicted by SuperdropNet, a neural network trained to predict 25 time steps into the future.
Bottom panels, blue -- Superdroplet-derived bulk moments vs. predictions on time step ahead by SuperdropNet. Results are shown only for initial conditions in the held-out testing data. Mean absolute percentage errors (MAPE) are shown for each comparison.
}
\label{scatterplot25}
\end{figure}

\subsection{Accuracy of Multistep Training}

Given that a network trained `offline' ($k=1$, eq. \ref{eq:loss}) could accurately predict the next time step, but not long-term evolution of moments, we increased rollout length during training to a maximum value of $k=25$. In general, we expected to observe an improvement in performance for predictions many time steps in the future, along with at least some degradation for single time step predictions as there will no longer be the only contributing factor our loss function. In fact, MAPE for tendencies
over single time steps decreased for all moments except for ${N_r}$, for which it increased from 171.17 to 242.61 (Fig. \ref{scatterplot25}-top panel). This did not strongly impact prediction errors for bulk moments one time step ahead (Fig. \ref{scatterplot25}-bottom panel). MAPE for bulk moments increased slightly with $k=25$ for $N_c$ (0.81 to 0.87) and $N_r$ (0.39 to 0.49) compared to the network trained with $k=1$ (Fig. \ref{scatterplot1}). For the mass moments, MAPE decreased from 2.89 to 1.95 when the rollout length was increased to 25. Given that multistep training did not lead to major or consistent degradation of the results for single time step predictions, we next examined its effects on longer rollouts.

\begin{figure}[h]
\centering
\includegraphics[width=13cm,height=5cm,scale=0.5]{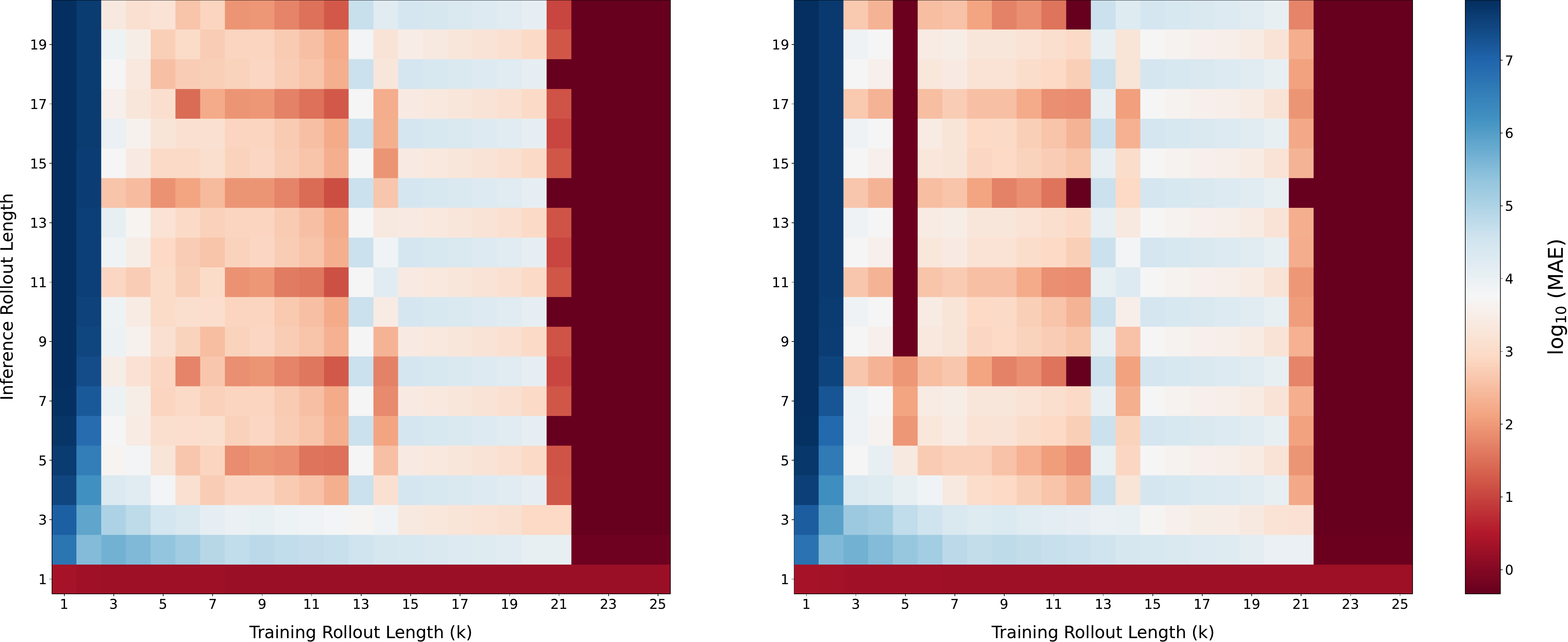}

\caption{Log$_{10}$ MAEs corresponding to model training steps and inference steps. The left panel includes all 719 training simulations and the right panel includes all 100 testing simulations.}

\label{error_plots}
\end{figure}

\begin{figure}[h]
\centering
\includegraphics[scale=0.5,trim={3cm 0 0 0.5cm},clip]{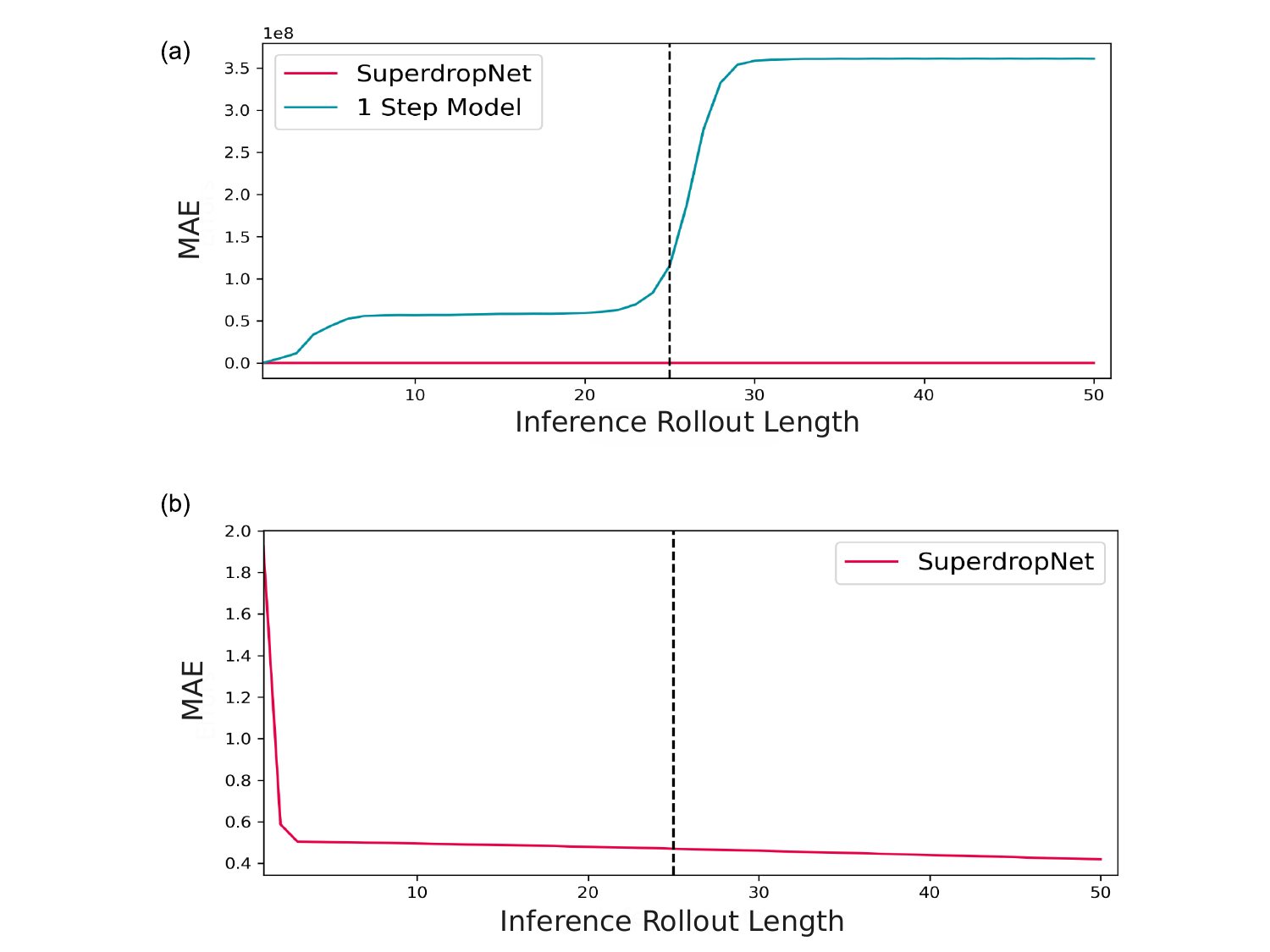}
\caption{Time evolution of errors as a function of rollout length during inference. (a) Mean absolute error as a function of rollout length using model trained to predict only one time step ahead (blue) vs. SuperdropNet. Errors are computed by averaging over 100 rollouts starting from randomly selected simulations and time points in the test set. The dashed black line shows SuperdropNet's training rollout length ($k=25)$. (b) As in `a,' but showing only SuperdropNet.}
\label{fig:errorvstime}
\end{figure}

\subsection{Rollout Lengths in Training and Evaluation}
The optimal rollout length for the training was not obvious: too low a $k$ value could reduce accuracy for longer prediction horizons, while too high could make optimization unstable or reduce accuracy for shorter horizons. To investigate this further, we calculated accuracy as a function of both $k$ and the prediction horizon (Fig. \ref{error_plots}). We calculated the mean MAE over normalized values of the four moments for all forecast horizons, from 1 to 20 steps, both on training/validation data (Fig. \ref{error_plots}-left panel) and on held-out testing data (Fig. \ref{error_plots}-right panel). 

For $10\leq k \leq 20$ changes were smaller and less consistent, while for $21\leq k \leq 25$ we again observed reduction of errors by several orders of magnitude for all prediction horizons with the exception of single time steps. This reduction of errors for prediction horizons longer than the training rollout length $k$ was only observed when using the pushforward trick ($\lambda_j=0, \forall j<k$, sec. \ref{ssec:multistep}), and not when setting all $\lambda_j=1$. Surprisingly, for $k>21$ we achieved better predictions for 2 or more time steps into the future than for single time steps, suggesting the network is correcting its own errors. 
Overall, accuracy on training/validation data strongly resembled accuracy on test data, suggesting that overfitting is negligible for this combination of network architecture, data and training procedure, and that our trained network can generalize to initial conditions not observed during training. In light of these results, we used the network trained with $k=25$ for all subsequent analyses, and termed it SuperdropNet. The network's ability to correct it's own errors can be seen clearly in Fig. \ref{fig:errorvstime}. While the prediction errors for the 1-step trained network (blue line) accumulate with longer rollouts, SuperdropNet's errors decrease over time (Fig. \ref{fig:errorvstime}-b).

\subsection{Rollout Accuracy Depends on Droplet Distribution Shape and Water Content}
We further investigated how SuperdropNet's accuracy over long rollouts depended on the conditions of the simulated warm rain process. Fig. \ref{ex_1} -\ref{ex_3} show SuperdropNet rollouts (dark-blue solid lines) over the full length of 3 simulations, starting from initial conditions. We also show rollouts for PRNet, which is the ML model developed to predict the process rates as in \cite{seifert_potential_2020}.  SuperdropNet produced stable output in each case which match the superdroplet simulation (black lines) better than the 1-step network (red-dashed lines), but with varying accuracy. 

Fig. \ref{ex_1} shows a scenario with high water content, in which SuperdropNet predictions match the superdroplet simulations better than the bulk moment scheme (dotted blue lines) while PRNet (yellow-dotted lines) converts clouds to rain faster than in superdroplet simulations. In Fig. \ref{ex_2}, a simulation with low water content and a higher shape parameter, SuperdropNet predictions show a significant improvement over the 1-step network's predictions. SuperdropNet and the 1-step network convert the cloud droplets to rain faster than superdroplet simulations, while the bulk moment scheme is a more accurate match. However, the bulk moment scheme overestimates rain droplet concentrations, while SuperdropNet does not.  In Fig. \ref{ex_3}, the 1-step network matches the superdroplet simulations better than SuperdropNet for the mass moments but SuperdropNet is a better match for the droplet concentrations. While the bulk moment scheme converts the cloud water to rain water faster than the superdroplet simulations, the bulk moment scheme takes longer than the superdroplet simulations. This is a case with relatively higher water content and a low value of shape parameter. PRNet overestimates the conversion time of cloud to rain in Fig. \ref{ex_2} and \ref{ex_3} . 

We observe a general pattern of SuperdropNet struggling with cases where the initial water content is low and the distribution of droplet sizes is wider (higher values of $\nu$). The PRNet model \cite{seifert_potential_2020} also struggled on simulations with a low water content. In general the bulk moment scheme closely predicts $L_{c}$ and $L_{r}$ but struggles with $N_{r}$.

\begin{figure}[h]
\centering
\includegraphics[scale=0.5,trim={0.7cm 0 0 0},clip]{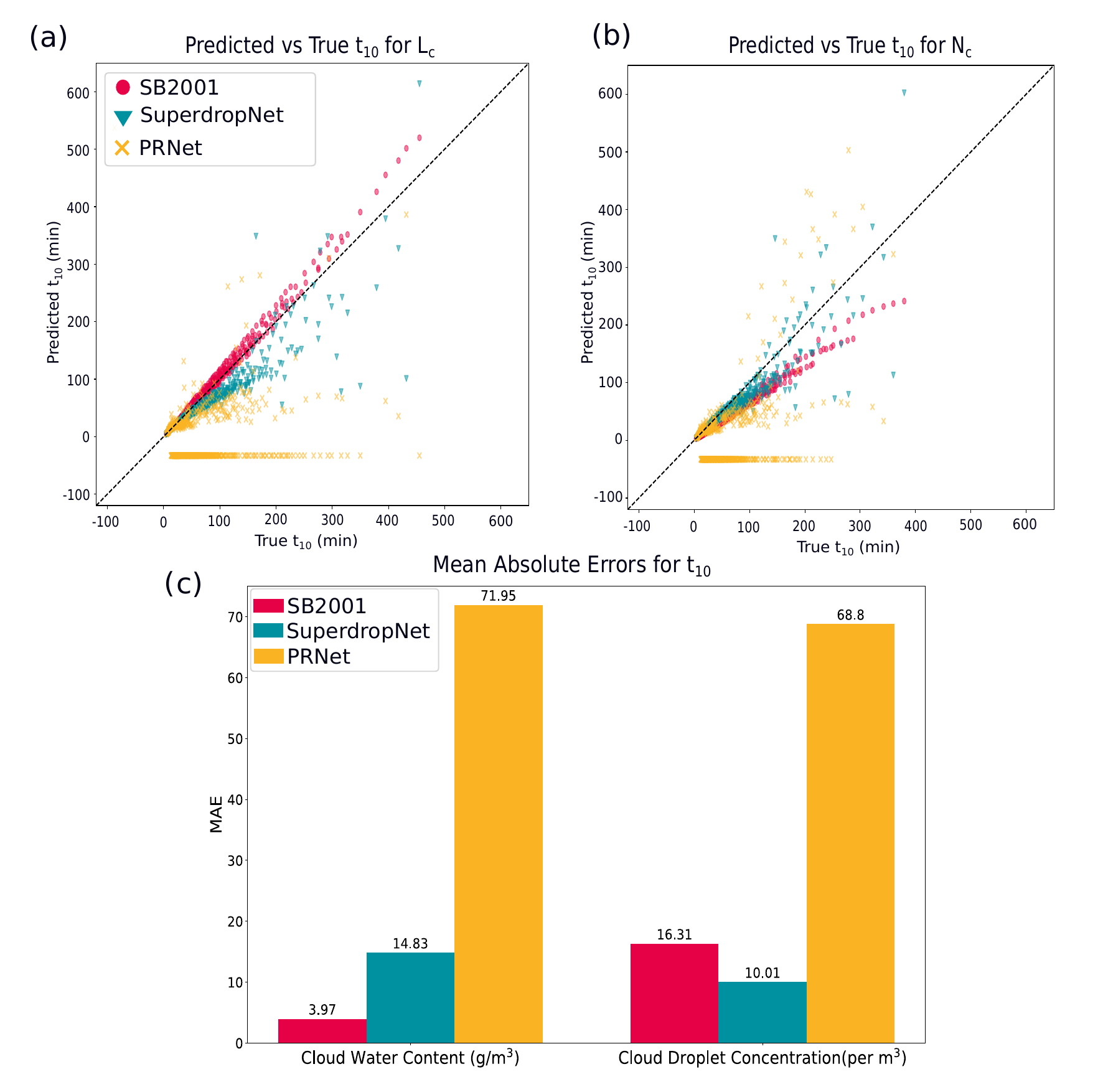}

\caption{The $t_{10}$ values for $L_c$ (a) and (b) $N_c$ as calculated for 819 simulations using SB2001, SuperdropNet and the ML model from \cite{seifert_potential_2020} based on estimating process rates, referred to as PRNet. Each point is one of the 819 simulations. True $t_{10}$ on the x axis corresponds to the values from the superdroplet simulations. The negative values on the y-axis represent simulations for which the 10\% of initial water never converted to rain during the length of the simulation. (c) Mean Absolute Errors (MAE) in approximation of $t_{10}$ values for cloud water content ($L_c$) and cloud number Concentration ($N_c$)}
\label{t10 plots with prnets}

\end{figure}

\begin{figure}[h]
\centering
\includegraphics[width=15cm,scale=0.5,trim={4cm 1.5cm 3cm 0cm},clip]{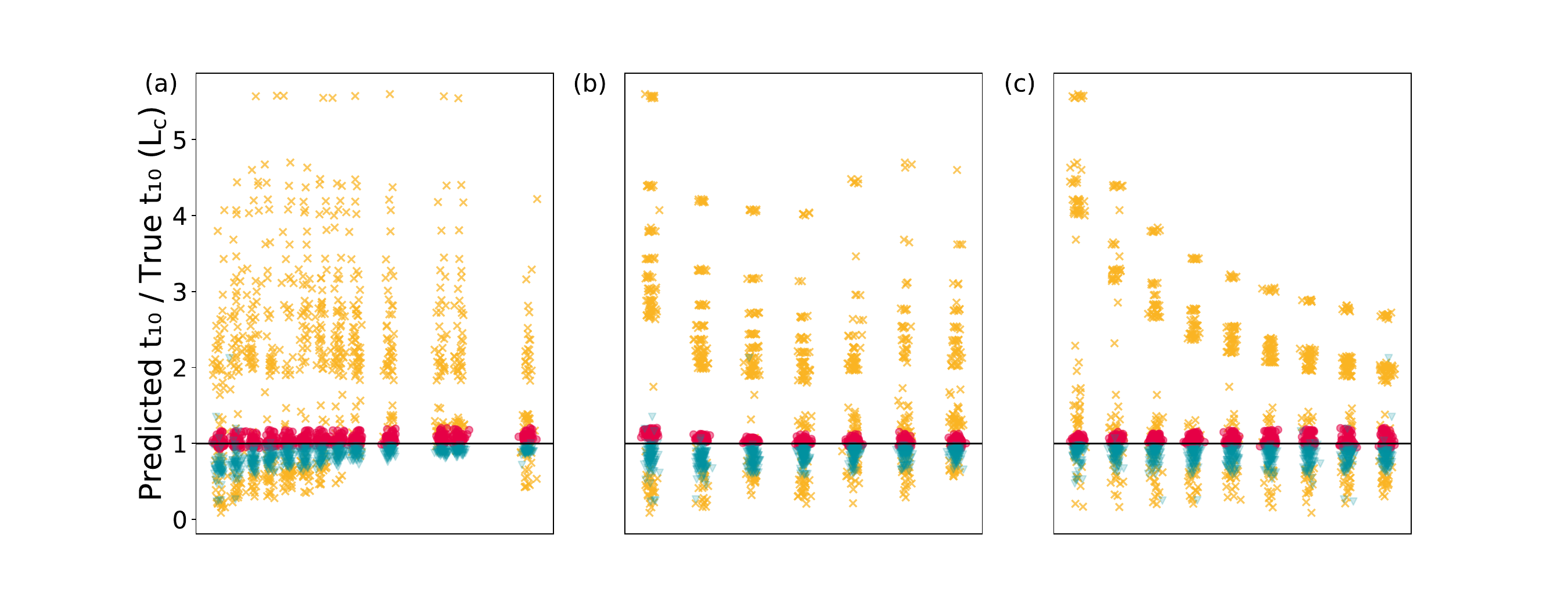}

\includegraphics[width=15cm,scale=0.5,trim={4cm 0 3cm 2cm},clip]{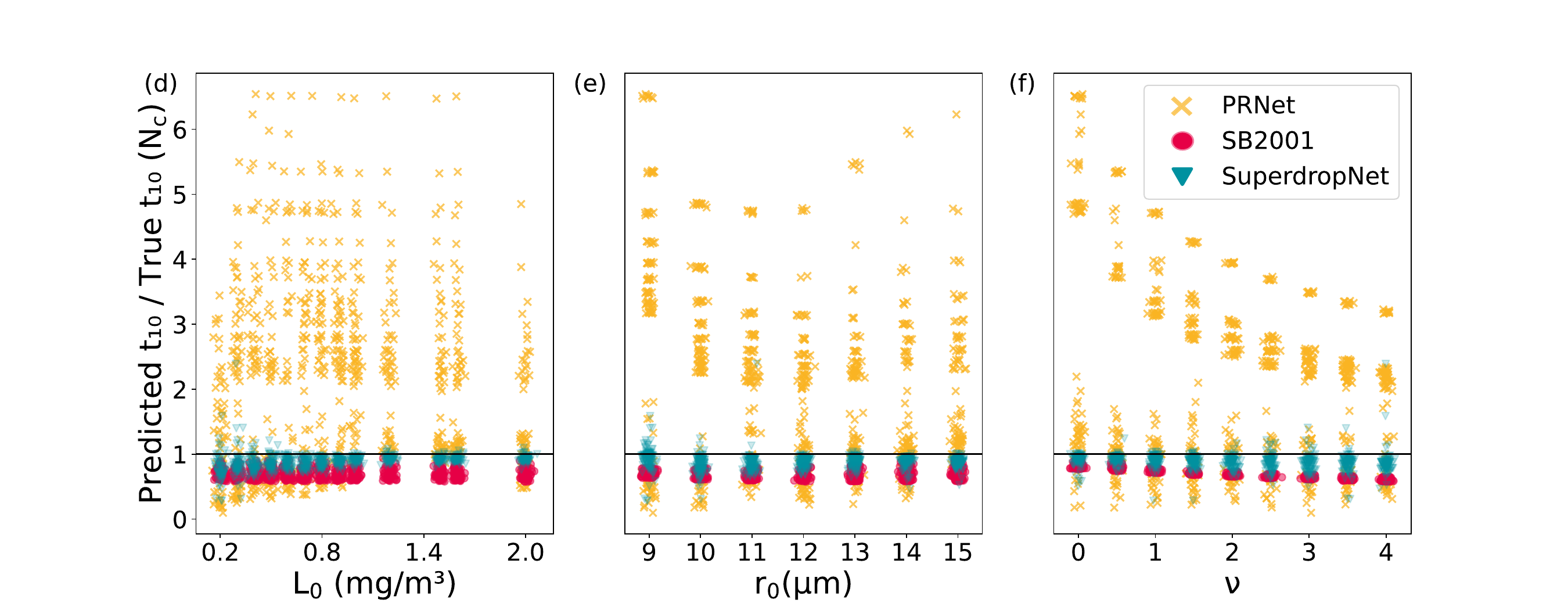}

\caption{(a)-(c) Ratio of predicted/true $t_{10}$ values for $L_c$ as a function of initial conditions of the superdroplet simulation. Here `true' refers to the superdroplet simulation. Each point is one of the 819 simulations. (d)-(f) same as (a)-(c) but for $t_{10}$ predictions for $N_c$. }

\label{t10 with ic}
\end{figure}

\subsection{Comparison of $t_{10}$ and Mean Absolute Errors}
An important test of any representation of warm rain coalescence is whether it accurately captures the timing of cloud-to-rain transitions. We therefore evaluated the match between SuperdropNet and alternative methods in the time $t_{10}$ at which 10\% of total water mass had converted to rain. In order to obtain a full picture over all initial conditions, and as previous analyses showed a lack of overfitting (Fig. \ref{error_plots}), we conducted this analysis on all 819 simulations. 

True $t_{10}$ values showed overall agreement with $t_{10}$ values computed from SuperdropNet rollouts (Fig. \ref{t10 plots with prnets}, blue, MAE=14.83 g/m$^3$) but tended to underestimate transition times. The classical bulk moment scheme exhibited a lower MAE (Fig.\ref{t10 plots with prnets}-(a), red, MAE = 3.97 g/m$^3$) but tended to overestimate transition times. PRNet underestimated many $t_{10}$ values to a greater extent than SuperdropNet (Fig. \ref{t10 plots with prnets}, yellow, MAE=71.95 g/m$^3$), but also frequently failed to reach 10\% mass conversion before the end of the simulation (shown as negative values). For these simulations, the MAE was calculated by assuming the end of the simulation as the $t_{10}$ value. 

We further examined the accuracy with which the timing of droplet number dynamics were represented, by calculating the time $t_{10}$ at which cloud droplet count had decreased by 10\% of its initial value. Here we observed a closer match to superdroplet simulation for SuperdropNet than for the bulk moment parameterization (Fig. \ref{t10 plots with prnets}-(b)). We also observe an overall lower MAE for SuperdropNet than the bulk moment scheme and PRNet (Fig. \ref{t10 plots with prnets}-(c)).

Given our finding that the accuracy of all schemes` bulk moment predictions depended on the simulation's initial conditions, we further examined how these initial conditions impacted the accuracy of $t_{10}$ values based on $L_c$ and $N_c$ (Fig. \ref{t10 with ic}). Consistent with our previous findings, SuperdropNet and PRnet performed best at higher water content. SuperdropNet's predictions also improved at higher initial droplet sizes (Fig. \ref{t10 with ic}-(b),(e)) and lower values of the shape parameter (Fig. \ref{t10 with ic}-(c-f)). For PRNet, the shape parameter affected the accuracy of prediction the most with an increased accuracy at higher values of ${\nu}$ (Fig. \ref{t10 with ic}-(c-f)). The bulk moment scheme's timing accuracy was largely unaffected by ${L_0}$ and ${r_0}$, but higher values of ${\nu}$ increased the magnitude of errors (Fig. \ref{t10 with ic}-(c-f)). SuperdropNet yielded more accurate transition times for ${N_c}$ across all initial conditions compared to the bulk moment scheme. 

Together, these results show that SuperdropNet improves significantly upon the state of the art in ML-based parameterization of warm rain microphysics. SuperdropNet closes most of the existing performance gap between previous ML-based representations and classical bulk moment schemes, and exceeds the accuracy of those classical schemes in some cases.

\section{Discussion}
We developed SuperdropNet, a neural network emulator of the warm rain formation process, and trained it on data from stochastic box-model simulations. SuperdropNet significantly advances the state of the art for ML-based emulation of warm-rain processes. For predicting the timing of cloud-to-rain transitions in droplet-based simulations, it closed most of the performance gap between neural networks and classical bulk moment parameterizations, and in some cases exceeded the accuracy of classical approaches. This performance gain can be attributed to several novel aspects of our approach: multistep rollouts, the pushforward trick, measurement and averaging out of stochasticity, and discarding the approximations used to link process rates for mass and droplet count. The striking improvements we observed with multistep training (Fig. \ref{error_plots}) can be considered a case of properly aligning our objective function with the ultimate task we wish to accomplish: emulating the warm rain dynamics accurately over many time steps.

We found SuperdropNet's accuracy to depend strongly on the conditions of the warm rain process, with better performance for high water content and a less-dispersed droplet distribution. In general these conditions make for an easier prediction task, as they lead to shorter but smoother cloud-to-rain transitions. On the whole, we found that SuperdropNet tends to underestimate  transition times calculated using rain vs. cloud mass, but predicts transitions of droplet counts more accurately than a classical bulk moment scheme.

Particle-based schemes allow for better approximation to the stochastic collection equation by estimating pure stochastic growth \cite{unterstrasser_collectionaggregation_2017}. These properties can be leveraged to improve representation of processes such as sedimentation where the bulk moment schemes are known to introduce numerical discontinuities \cite{wacker_evolution_2001,seifert_two-moment_2006}. Additionally, initial successes in a warm rain scenario with only rain and cloud open the possibility of emulating more complex microphysical phenomena involving additional hydrometeors such as ice, snow and graupel, as proposed in a previous study \cite{seifert_potential_2020}.

We further plan to couple SuperdropNet to atmospheric fluid dynamics and other physical processes. This will require solving the technical challenge of bidirectional communication between Python/Pytorch-based deep learning and FORTRAN based atmospheric simulation, as well as developing new training algorithms that encourage stability and accuracy of the coupled dynamics. We believe that fast, accurate microphysics emulators that can be used as modular simulation components could offer significant benefits in predicting and understanding weather, climate and air quality phenomena.

\section*{Acknowledgements}
We thank Axel Seifert for providing access to McSnow which was used for generating the training data. This work was funded by Helmholtz Association's Initiative and Networking Fund through Helmholtz AI. We also thank Ann-Kristin Naumann for insightful scientific discussions.

\clearpage
\bibliographystyle{unsrt}

\end{document}